\documentclass[allclo]{FBSart}
\usepackage{epsfig,psfig}
\newcommand{\be}{\begin{equation}}
\newcommand{\ee}{\end{equation}}
\newcommand{\ba}{\begin{eqnarray}}
\newcommand{\ea}{\end{eqnarray}}

\def\vec#1{{\mbox{\boldmath$#1$}}}
\def\ket#1{\vert #1 \rangle}
\def\bra#1{\langle #1 \vert}

\title{
\[ \vspace{-2cm} \]
\noindent\hfill\hbox to 1.5in{\rm  } \vskip 1pt
\noindent\hfill\hbox to 1.5in{\rm SLAC-PUB-10779 \hfill  } \vskip
1pt \noindent\hfill\hbox to 1.5in{\rm June 2, 2004 \hfill}\vskip
10pt CORE: Frustrated Magnets, Charge Fractionalization and QCD\footnote{Presented
at Light-Cone 2004, Amsterdam, 16 - 20 August}\footnote{This work was
supported by the U.~S.~DOE, Contract No.~DE-AC03-76SF00515.}}
\author{Marvin Weinstein}
\institute{Stanford Linear Accelerator Center, Stanford University,
  Stanford, California 94309}
\runningauthor{Marvin Weinstein}
\runningtitle{LC 2004}
\begin{document}
\maketitle
\begin{abstract}
I explain how to use a simple method to extract the physics of
lattice Hamiltonian systems which are not easily analyzed by exact or
other numerical methods.  I will then use this method to establish the
relationship between QCD and a special class of generalized, highly frustrated
anti-ferromagnets.
\end{abstract}
\section{Introduction}

The title of my talk "CORE: Frustrated Magnets, Charge
Fractionalization and QCD" might seem peculiar for a conference on
light-front physics, however I hope to convince you that my topic
is more relevant than it appears to be. Given my background,
the examples I discuss won't be in the light-front formalism but
I think it will be obvious that the general technique could be
fruitfully adapted to treat the problems Simon Dalley talks about
when he discusses the transverse lattice.

Since CORE\cite{COREpaper} is a method for analyzing the physics
of a general class of Hamiltonian lattice field theories it is
clearly relevant to this class of problem. The question,
"Why am I talking about frustrated
magnets and charge fractionalization?", requires a longer
explanation. I must begin by telling you what frustrated
magnets are, and then I can tell you what makes them interesting.
Finally, I have to tell you why I am talking about charge
fractionalization (why QCD should be obvious).

\section{Frustrated Magnets}

The $1+1$-dimensional Heisenberg anti-ferromagnet (HAF) is
defined by Hamiltonian
\be
    H = \sum_j \vec{s}(j)\cdot \vec{s}(j+1) .
\label{HAF}
\ee
The variable $j$ in Eq.\ref{HAF} is an integer labelling the sites
of a one-dimensional spatial lattice and I assume that
$-\infty < j < \infty$. The operator $\vec{s}(j) = 1/2
\,\vec{\sigma}(j)$, where $\vec{\sigma}$ is a Pauli spin matrix,
and it acts on the spin-$1/2$ degree of freedom associated with
each lattice site $j$.

Intuitively, a good candidate for a mean-field approximation to
the ground state of this system is obtained by having the average
value of the spin operator $\bra{\psi}\vec{s}(j) \ket{\psi}$
reverse direction moving from site to site; i.e., this Hamiltonian
favors states in which neighboring spins anti-align.

"What is a frustrated anti-ferromagnet?". The simplest
example is one where the frustration is geometrical in
origin.  For example, consider generalizing our
one-dimensional HAF to a two-dimensional triangular
lattice. Focusing on any one triangle we see that, if the
spins on any two of the sites of the triangle anti-align,
the third spin doesn't know what to do. Frustration can also
arise when there are long range couplings.  For example,
if we add next-to-nearest neighbor couplings to an HAF; i.e.,
\be
    H = \sum_j \left[ \alpha_1\,\vec{s}(j)\cdot \vec{s}(j+1) + \alpha_2\,\vec{s}(j)\cdot
    \vec{s}(j+2) \right] ,
\label{frustHAF}
\ee
then, when $\alpha_1 = \alpha_2$, there is once again no
way to get a low-energy pattern of anti-aligned spins.

Frustrated anti-ferromagnets are interesting for several reasons.
First, although they are easily defined, they are not well
understood. Semi-classical computations suggest that the
example defined in Eq.\ref{frustHAF} has a devil's staircase of
phase transitions; i.e., as the ratio $\alpha_1/\alpha_2$
approaches a fixed value the system undergoes an infinite number
of phase transitions.  Another reason is that it is conjectured
they exhibit charge fractionalization in all dimensions. This
phenomenon is known to occur in $1+1$-dimensional theories, but is
not known to occur in system with more spatial dimensions.
Basically, charge fractionalization means that what appears, at
the level of microphysics, to be a theory of neutral bosons, turns
out, at low energy, to be best describe in terms of a theory of
charged fermions.

\section{CORE Basics}

As I already said, CORE is an acronym for the COntractor REnormalization group
technique. It can be used to extract the physics of Hamiltonian
lattice systems which are amenable neither to exact solution, nor
conventional numerical techniques. A cartoon of the
procedure followed in a CORE computation is shown in
Fig.\ref{Fig1}, where I specialize to the case of a $1$-dimensional
spatial lattice, where a spin-$1/2$ degree of freedom is associated
to each lattice site $j$.  My point is to show how to
define a procedure for truncating the original Hilbert space to
an appropriately chosen subspace and then, how to construct a new Hamiltonian,
defined on this subspace, which has exactly the same low-energy
physics as the original theory.

To construct the truncation of the Hilbert space
we divide the lattice into finite size blocks.  In the
cartoon each of these blocks is assumed to contain three sites.
Associated with site in a block there is a spin-$1/2$ degree of freedom
so each block represents eight possible states.  We next restrict the total
Hamiltonian to a single block, which leads to an $8\times 8$ matrix which is
easily diagonalized.  From its lowest lying eigenstates we then select a subset
called the {\it retained states\/}, and define
the {\it restricted Hilbert space\/} to be the one
constructed by taking all tensor products of the retained
states.  Thus, for example, retaining the two lowest lying states
reduces a problem on a lattice with $V$ sites and an original
$2^V$-dimensional Hilbert space to a new lattice with
$V/3$ sites and a $2^{V/3}$-dimensional space.

Using the projection operator corresponding to the construction
defined above, one computes the {\it renormalized\/} or {\it
effective\/} Hamiltonian, $\tau(H)$, by evaluating the following
formulas.
\ba
    P(j) &=& \ket{\phi_s(j)}\bra{\phi_s(j)} \quad P = \prod_j P(j), \label{projdef}\\
    \tau(H) &=& \lim_{t\rightarrow \infty}
            [[T(t)^2]]^{-1/2}\, [[T(t) H T(t)]] \, [[T(t)^2]]^{-1/2},
            \label{tauH}
\ea
where $T(t) = e^{-t H}$ and where $[[O]]= P O P$, for any
any operator $O$.

Obviously, exactly evaluating this formula,
which involves the exponential of the Hamiltonian, requires that one
already knows all of the eigenvalues and eigenstates of the
original Hamiltonian. Of course, if we know that, there is no point to
this exercise.  The CORE method is useful because there is a way to
compute the renormalized Hamiltonian, to arbitrary accuracy, without
being able to solve the problem exactly.  This can be done because
the renormalized Hamiltonian can always be written
as a sum of finite-range connected operators; i.e.,
\be
            \tau(H) = \sum_j \sum_r h_r(B_j,B_{j+1},\ldots,B_{j+r-1}).
\label{clusterform}
\ee
Here, each {\it connected range-$r$\/} operator
$h_r(B_j,B_{j+1},\ldots,B_{j+r-1})$ is a product of operators
which act only on the spins in $r$-adjacent blocks. (A similar
formula can be written to define the {\it renormalized\/} version
of any other extensive operator.)
It is remarkable that an accurate computation of the coefficient of
each term in this expansion can be achieved by working on small
sublattices.  Furthermore, it is also true that one can get
a very good approximation to the full renormalized Hamiltonian
keeping only a few terms in the cluster expansion, say to range
four or five.

The first approximation to the coefficient of a range-$r$
connected operator comes from evaluating Eq.\ref{tauH} for
a connected sublattice containing $r$-adjacent blocks and then
subtracting from the result those pieces of $\tau(H)$ already
computed in the previous computations for smaller $r$.  This coefficient
is quickly improved by computing the higher range computations.
The formulae for the range-$1$ and range-$2$ connected operators are
\ba
    h_1{B_j} &=&  \tau(H(B_j)) \\
    h_2(B_j,B_{j+1}) &=& \tau(H(B_j,B_{j+1}) - h_1(B_j) - h_1(B_{j+1}).
\ea
The formulae for larger $r$ are obtained by generalizing this procedure.

\subsection{Example 1: Heisenberg Anti-Ferromagnet}

\begin{figure}[3in]
\begin{center}
\psfig{figure=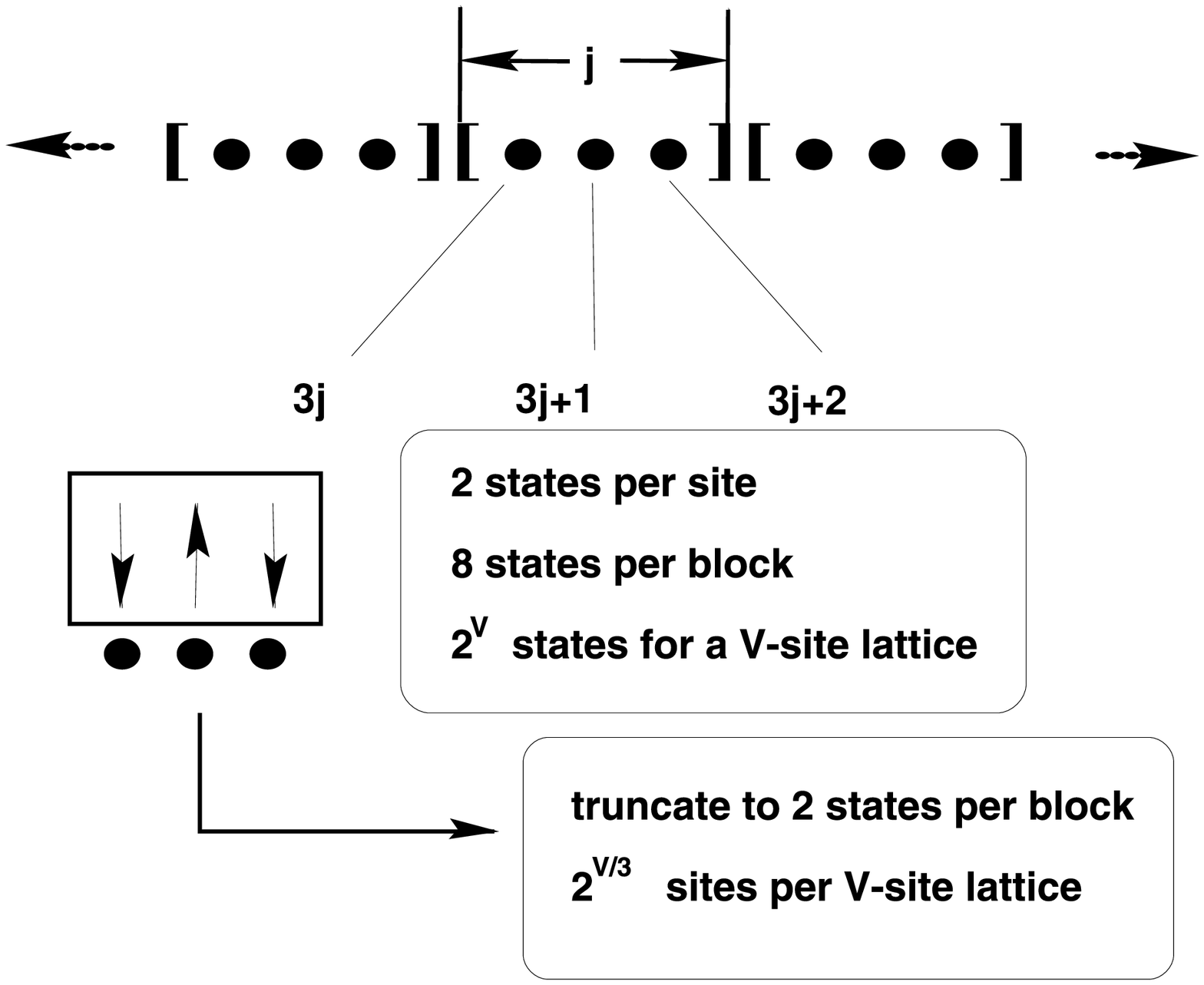,height=3in}
\caption{\label{Fig1} Subdivision of $1$-dimensional lattice and truncation scheme.
}
\end{center}
\end{figure}

Let's skip further generalities and see how this works for the
case of the HAF defined in Eq.\ref{HAF}.
As in the cartoon in Fig.\ref{Fig1}, divide the lattice into $3$-site blocks.
In this case the single block Hamiltonian is
\ba
    H(B_j) &=&
    \left[ \vec{s}(3j)\cdot \vec{s}(3j+1)
    + \vec{s}(3j+1)\cdot \vec{s}(3j+2)\right]\\
&=& \vec{s}(3j+1)\cdot\left[\vec{s}(3j) + \vec{s}(3j+2)\right]\\
&=&  \frac{1}{2} \left[ S_{0+1+2}^2 - S_{0+2}^2-\frac{3}{4}\right]\label{ex3sites} ,
\ea
where by $S_{0+1+2}^2$ and $S_{0+2}^2$ we mean the sum of the
squares of the total spin operators for sites $3j,3j+1,3j+2$ and
$3j,3j+2$ respectively.  Clearly Eq.\ref{ex3sites} tells us that
the eigenstates of the $3$-site problem correspond to two
spin-$1/2$ multiplets and one spin-$3/2$ multiplet.  Since the
lowest lying spin-$1/2$ multiplet is the one for which
$S(0+1+2)^2=3/4$ and $S(0+2)^2=2$, we will use only this pair of
degenerate states to define the space of{\it retained states\/} .
Since these state have the same energy, it immediately follows that
\be
    h_1(B_j) = - {\bf 1}_j .
\ee

To compute the range-$2$ contribution, $h_2(B_j,B_{j+1})$,  we have to solve the $6$-site
problem exactly and evaluate $\tau(H)$ for the projection operator constructed from the
four states obtained by taking tensor products of the two lowest-lying spin-$1/2$ states in
each block.  This computation is made quite simple if we observe that these four states
can be recombined into one spin-$0$ and one spin-$1$ multiplet as follows:
\be
\ket{\uparrow \uparrow}, \quad \frac{1}{\sqrt{2}} \left( \ket{\uparrow \downarrow} ++
\ket{\downarrow \uparrow} \right) ,\quad \ket{\downarrow\downarrow}
\ee
\be
\frac{1}{\sqrt{2}} \left( \ket{\uparrow \downarrow} - \ket{\downarrow \uparrow}
\right)
\ee
Each of these states has a definite total spin and definite
total $z$-component of spin and since the total $6$-site
Hamiltonian commutes with these operators, it follows that $T(t)$
contracts each of these states onto the unique lowest energy
$6$-site state having the same quantum numbers.  Thus, in the total spin
basis $\tau(H)$ will have the following diagonal form, where $E_0$
is the energy of the lowest spin-$0$ state and $E_1$ is the energy
of the lowest lying spin-$1$ multiplet.
\ba
\left(
\begin{array}{c c c c }
    E_0 & 0 & 0 & 0 \nonumber\\
    0 & E_1 & 0 & 0 \nonumber\\
    0 & 0 & E_1 & 0 \nonumber\\
    0 & 0 & 0 & E_1 \nonumber\\
\end{array} \right) .
\ea
The correct eigenvalues are
\be
 E-0 = -2.493577\ldots \quad;\quad E_1 = -2.001995\ldots  \quad.
\ee
Following the approach we just used it is clear that,
in the original basis of tensor products of retained states, this Hamiltonian
takes the form
\be
   \tau(H(B_j,B_{j+1}) = \sum_j \left[
   \alpha\,{\bf 1} + \beta\,\vec{s}(j)\cdot\vec{s}(j+1) \right] ,
\ee
where $ \alpha = (E_0+3\,E_1)/4 = -3.99507 $ and $\beta = E_1 - E_0 = .491582$.

Since, after one step the new effective Hamiltonian
is a multiple of the unit matrix plus $\beta$ times the
original Hamiltonian, we see that this HAF Hamiltonian, in
range-$2$ approximation, is at a fixed point of this
renormalization group transformation.  In other words, no matter how
many times we repeat this process we always get a
Hamiltonian of the form
\be
   \tau(H)_n= \sum_j \left[
   \alpha_n\,{\bf 1} + \beta^n\,\vec{s}(j)\cdot\vec{s}(j+1) \right] .
\ee
Moreover, since $\beta < 1$, we see that the interaction term in the
Hamiltonian eventually iterates to zero, which tells us that we
are dealing with a massless theory.

The observation that $\beta^n \rightarrow 0$ tells us that
all we have to do is compute the limiting value of $\alpha_n$
divided by the number of sites on the initial lattice to obtain
the ground-state energy density.  Now, since after the first renormalization
group transformation the lattice has $V/3$ sites, the total
effect of the term proportional to $\alpha$ is to contribute
$\alpha\,V/3$ to the energy of all states in the new effective
theory.  Thus, dividing by $V$ we obtain a contribution of
$\alpha/3$ to the ground state energy density.  At this point the
simplest thing to do is subtract the constant term
from the Hamiltonian and do another renormalization group
transformation.  This time the new constant term will be $\beta
\,\alpha$ and will correspond to a theory defined on a lattice
with $V/3^2$ sites and so its contribution to the energy density
will be $\beta\,\alpha/3^2$.  Proceeding in this manner we see
that the total energy density is given by the geometric series
\be
    {\cal E} = \frac{\alpha}{3} \sum_n \left(\frac{\beta}{3}\right)^n =
\frac{\alpha}{3\left(1-\beta/3\right)}.
\ee
Plugging in the values of $\alpha$ and $\beta$ obtained from the $6$-site computation
we arrive at the results
\be
 E_{\rm ren-group} = -0.448446\ldots \quad ; \quad E_{\rm exact} = -0.442147\ldots
\ee
which shows that the simple range-$2$ approximation is accurate to about $1\%$.
Not bad for a simple first principles computation based upon diagonalizing at most
$4\times 4$ matrices.

\subsection{Example 2: The $1+1$-dimensional Ising Model}

Now that I have shown you that even the crudest approximation to
the exact CORE transformation gives qualitatively good results for
the Heisenberg anti-ferromagnet I wish to spend a few moments
showing you the kind of accuracy one can obtain by working a little harder. To
do this I consider the case of the $1+1$-dimensional Ising
model in a transverse field; i.e., the theory defined by the Hamiltonian
\be
    H = \sum_j -\left[ \cos(\lambda)\,\sigma_z(j) + \sin(\lambda)\,\sigma_x(j)\,
    \sigma_x(j+1) \right]
\ee

This theory is interesting for several reasons.  First, it clearly
undergoes a quantum phase transition as $\lambda$ varies from
$\lambda=0$ to $\lambda=\pi/2$. To see this observe that for
$\lambda = 0$ the theory has a unique ground state; namely, the
one which is a product of eigenstates of $\sigma_z(j)$
corresponding to the eigenvalue $+1$.  However, for $\lambda =
\pi/2$ there are two possible ground states made by taking the
product of eigenstates of $\sigma_x(j)$. The first, is a product
of eigenstates where the eigenvalue of $\sigma_x(j)$ are all equal
to $+1$; the second, where all the eigenvalues are equal to $-1$.
This tells us that as $\lambda$ varies the system goes from having
a unique ground state, to having a doubly degenerate ground state
corresponding to spontaneous breaking of the symmetry $\sigma_x(j)
\rightarrow -\sigma_x(j)$.  Even more interesting is the fact that
the low-lying excitations of the theory for small $\lambda$
correspond to having one spin flipped from up to down, but the
excitations in the case of $\lambda$ near $\pi/2$ are kinks or
anti-kinks (i.e., states where half of the system seems to be in
one ground state while the rest is in the other). I will now show
that a simple CORE computation is not only able to compute the
ground state energy density to high accuracy, but it correctly
finds the nature of the excitation spectrum and computes both the
mass gap and magnetization (i.e., the mean value of the operator
$(1/V)\,\sum \langle \sigma_x(j) \rangle$ as a function of
$\sin(\lambda)$.

\begin{figure}[2.5in]
\hbox{\hskip-.5in
\vtop to 3.25in{\hsize 3.25in
\psfig{figure=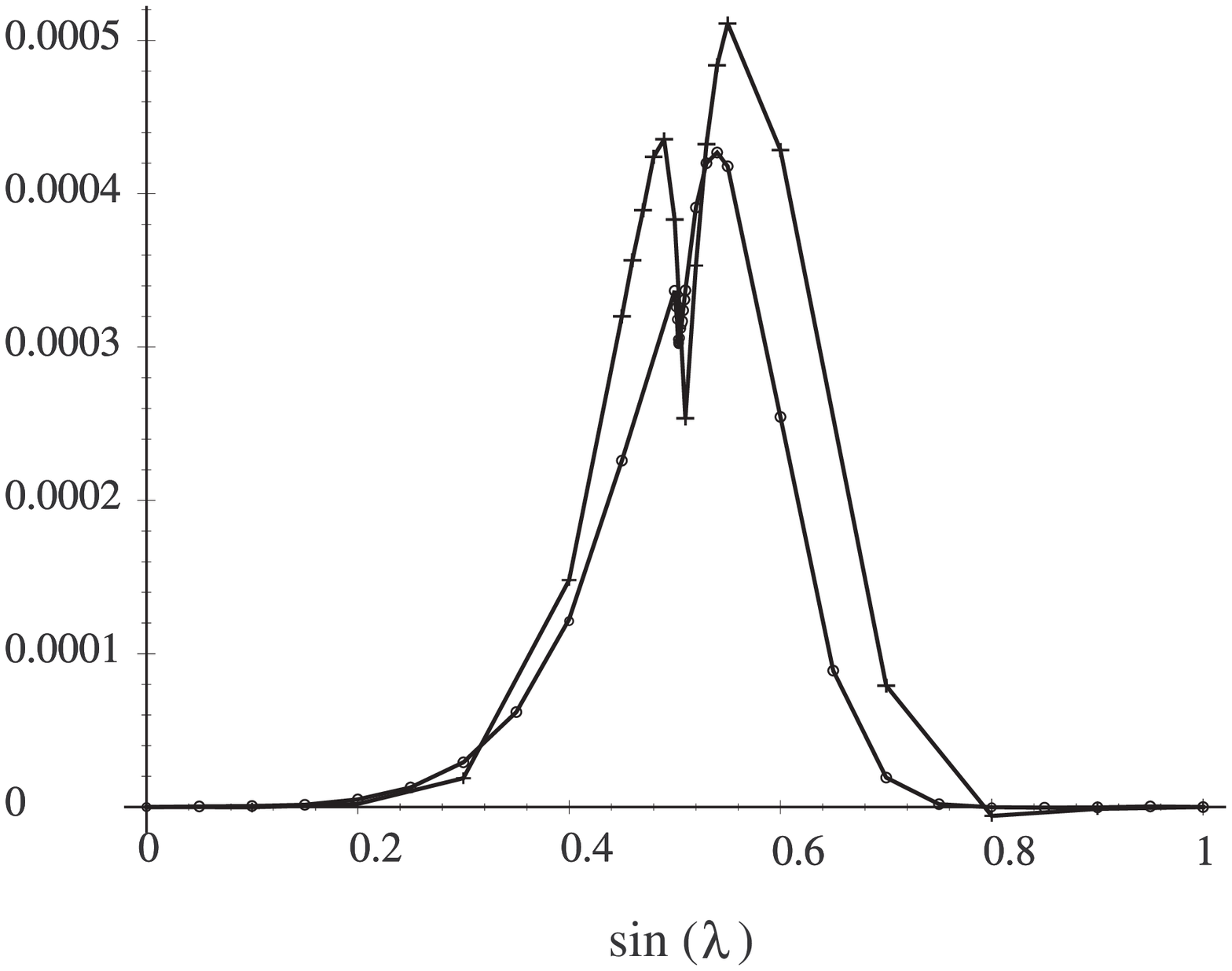,height=2.3in}
\caption{\label{Fig2} The difference between the exact ground state energy density
and the CORE results divided by the exact ground state energy density plotted
as a function of $\lambda$ }}\quad
\vtop to 3.25in{\hsize 3.2in
\psfig{figure=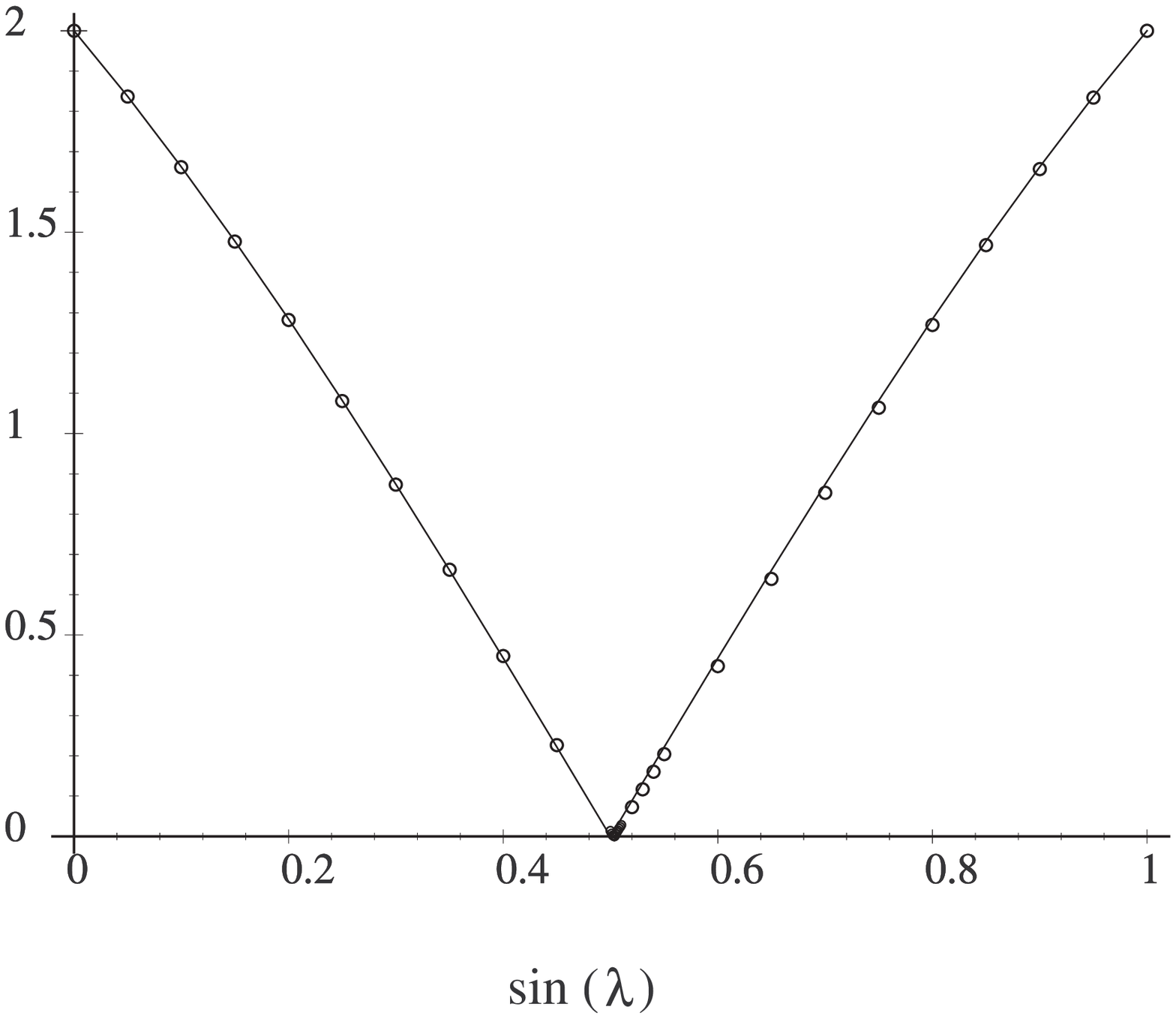,height=2.3in}
\caption{\label{Fig3} The exact mass-gap (solid curve) versus point-by-point
CORE computation (open circles) plotted as a
function of $\lambda$}}}
\end{figure}

\begin{figure}[2.5in]
\hbox{\hskip-.5in
\vtop to 3.25in{\hsize 3.25in
\psfig{figure=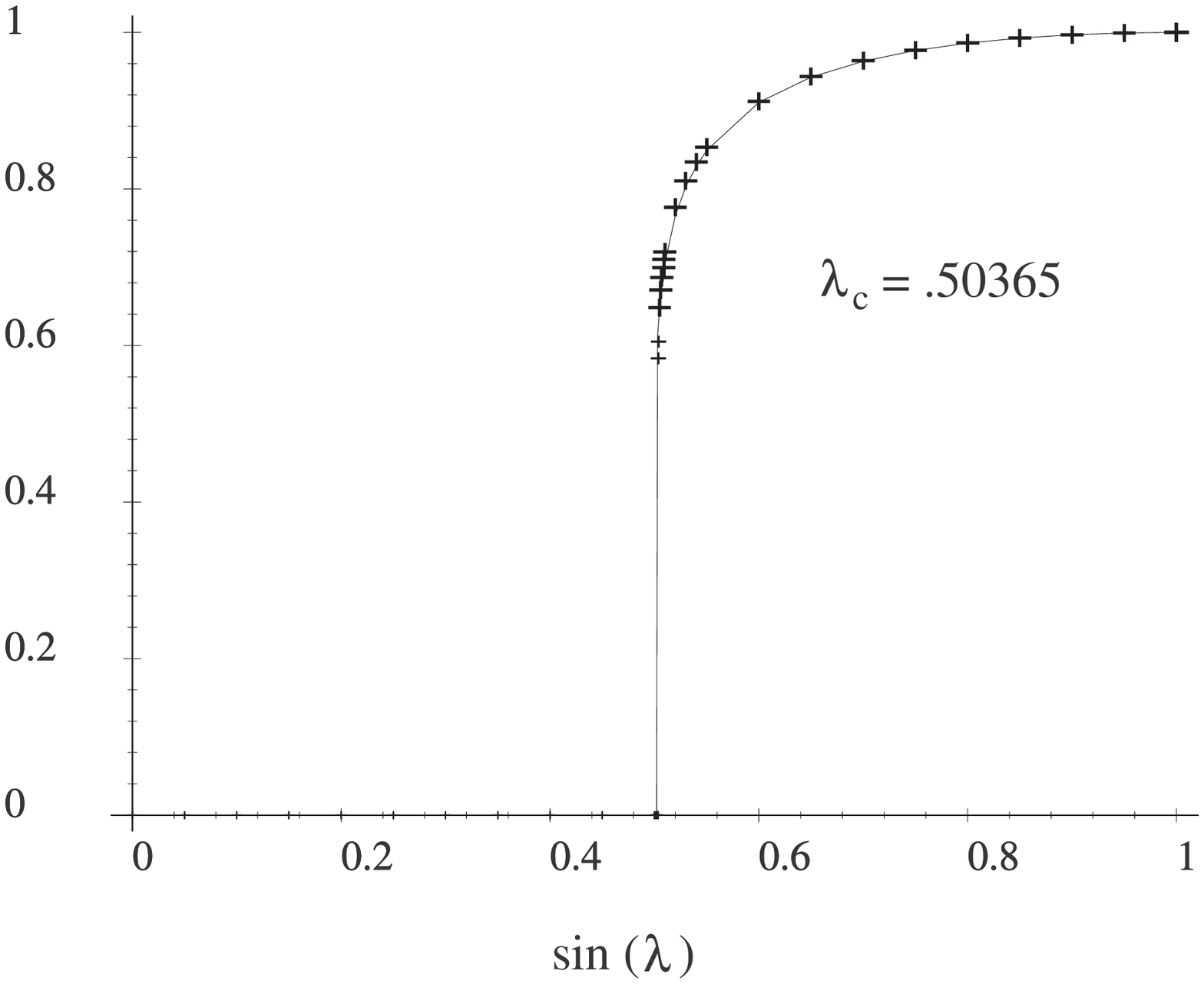,height=2.3in}
\caption{\label{Fig4} Exact magnetization versus (solid curve) versus point-by-point
CORE computation (open circles) plottedas a function of $\lambda$ }}\quad
\vtop to 3.25in{\hsize 3.2in
\psfig{figure=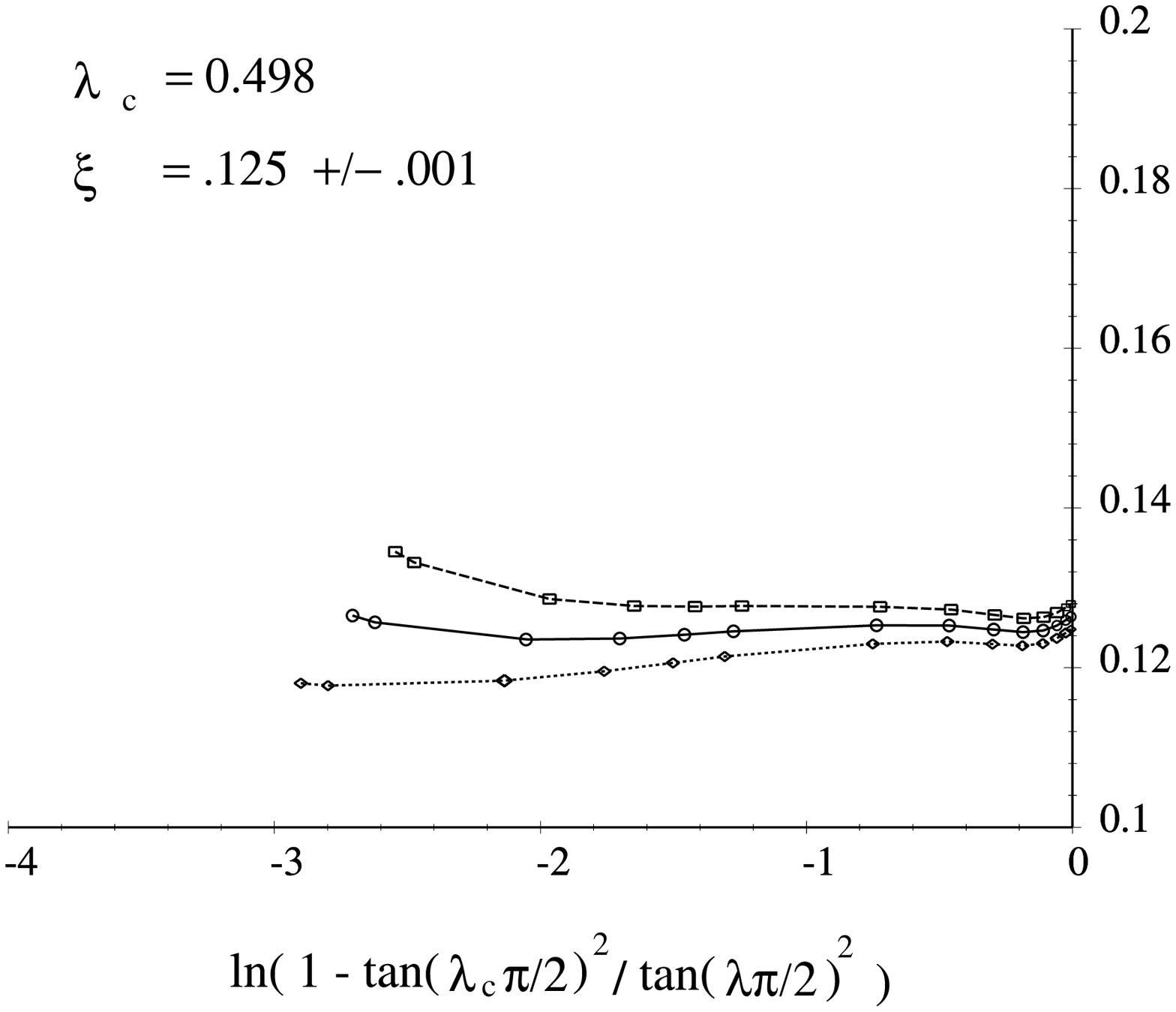,height=2.3in}
\caption{\label{Fig5} Extraction of the critical point and critical exponent for the
magnetization by fitting to the indicated form}}}
\end{figure}

Figs.\ref{Fig2}-\ref{Fig4} plot results obtained by carrying out
range-$3$ CORE computations for specific values of
$\sin(\lambda)$. Range-$3$ computations begin by
dividing the lattice into $3$-site blocks, and retaining the two
lowest lying eigenstates of the $3$-site Hamiltonian and then
solving the $6$-site and $9$-site problems exactly. Given
the accuracy of the results it is remarkable that the hardest
computation one does is to diagonalize a $512\times 512$ matrix.
Fig.\ref{Fig3} plots the difference between the
exact and CORE values for the ground state energy density divided
by the exact energy density for a range of values of
$\sin(\lambda)$.  This is done to make the errors visible.
The dotted curve corresponds to results obtained
for a range-$2$ ($6$-site) computation, whereas the solid curve gives the
results for a range-$3$ computation.  Clearly the method makes its
biggest errors near the phase transition, $\sin(\lambda)=.5$, but
even there it does well. In Fig.\ref{Fig4} we see a plot of
the exact mass gap (the solid curve) against the CORE
approximation to the mass gap (open circles).  The only
significant errors on this curve are near the critical point and
they are essentially due to the fact that at range-$3$ CORE makes
a small mistake in the location of the critical point.
Fig.\ref{Fig5} is a plot of the exact magnetization (solid curve)
above the phase transition versus the values computed by CORE on a
$\lambda$-by-$\lambda$ basis.  It is notoriously difficult to
compute this curve in a simple way since the behavior corresponds
to a critical exponent of $1/8$.  The final figure,
Fig.\ref{Fig6}, shows how one can extract both the critical point
and the exponent for the magnetization by varying both to produce
a straight line.

\section{Mapping One Theory To Another}

At his point I have shown how to use CORE to extract the physics of a
lattice Hamiltonian theory to high accuracy.  In the remainder of
this talk I will focus on what one can learn by using
CORE to map a theory into a very different looking but equivalent
Hamiltonian system.

\subsection{Example: Massless Bose Free Field}

Let us first consider the case of a massless Bose
free field in $1+1$-dimensions whose Hamiltonian is
\be
    H = \sum_j \left[ \frac{1}{2}p(j)^2 + \frac{\mu^2}{2}x(j)^2 + \left(
    x(j+1)-x(j)\right)^2 \right] .
\ee
My objective is to show that this theory can be mapped into an equivalent
spin system by truncating the single-site theory to the two lowest lying oscillator states.
Furthermore, I will show that one can obtain a very good approximation to the exact
renormalized Hamiltonian by including only a few terms in the cluster expansion.

\begin{figure}[2.5in]
\hbox{\hskip-.5in
\vtop to 3.25in{\hsize 3.25in
\hskip-.3in \psfig{figure=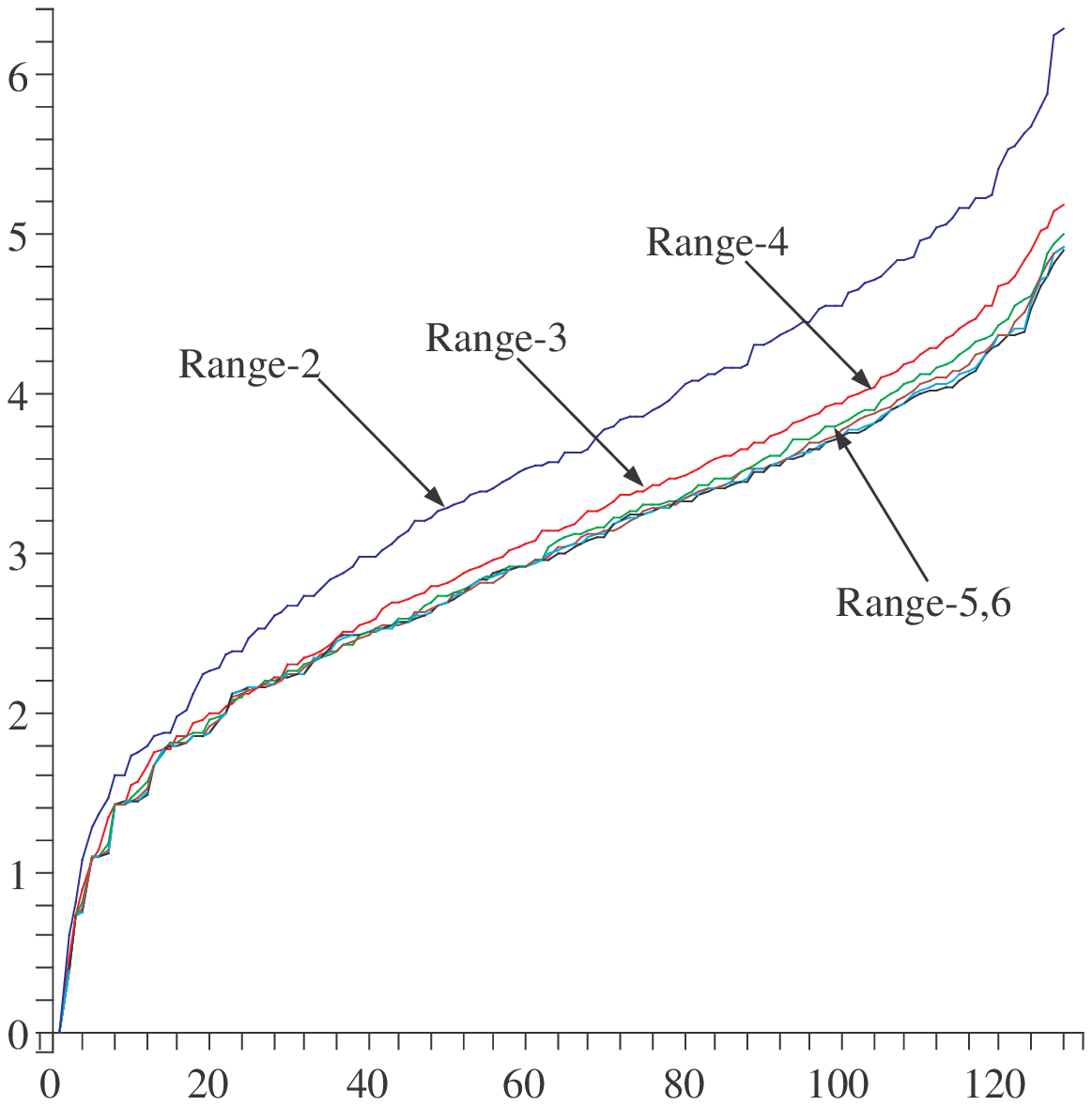,width=3.5in}
\caption{\label{Fig6} 6-Site Comparison}}\quad
\vtop to 3.25in{\hsize 3.25in
\hskip -0.3in
\psfig{figure=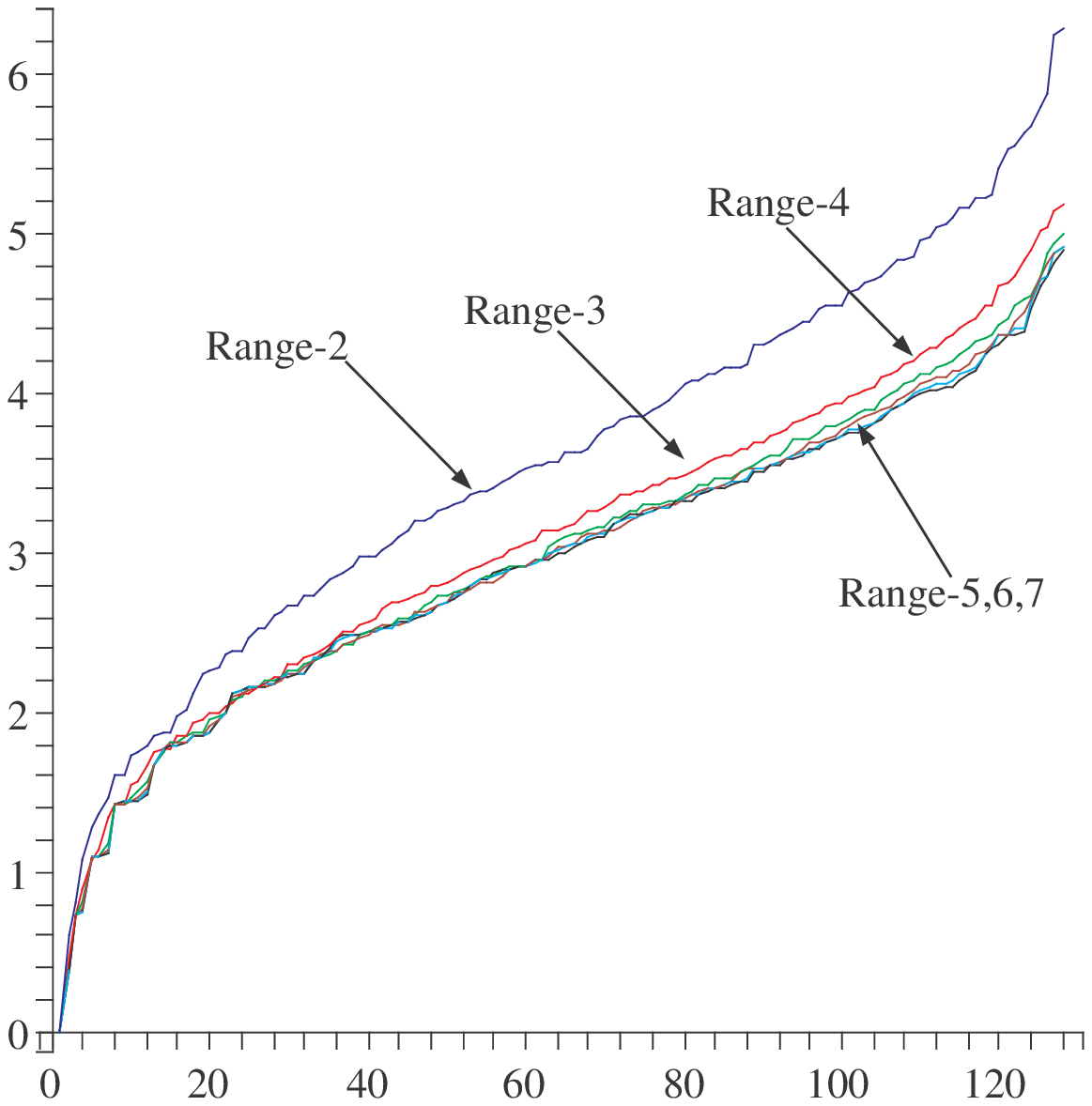,width=3.5in}
\caption{\label{Fig7} 7-Site Comparison} }}
\end{figure}

\begin{figure}[2.5in]
\hbox{\hskip-.5in
\vtop to 3.25in{\hsize 3.25in
\hskip-.3in \psfig{figure=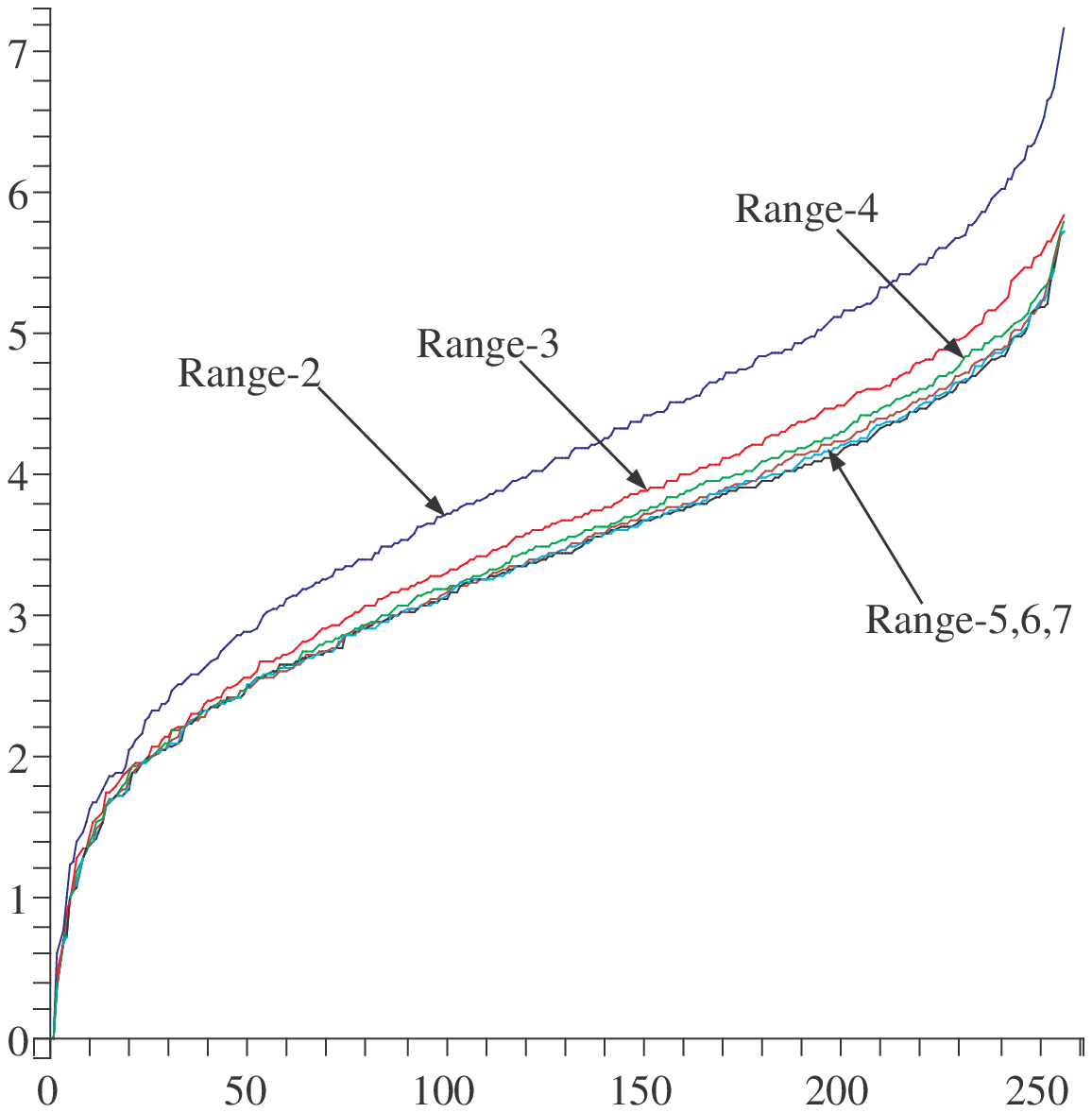,width=3.5in}
\caption{\label{Fig8} 8-Site Comparison }}\quad
\vtop to 3.25in{\hsize 3.25in
\hskip -0.3in \psfig{figure=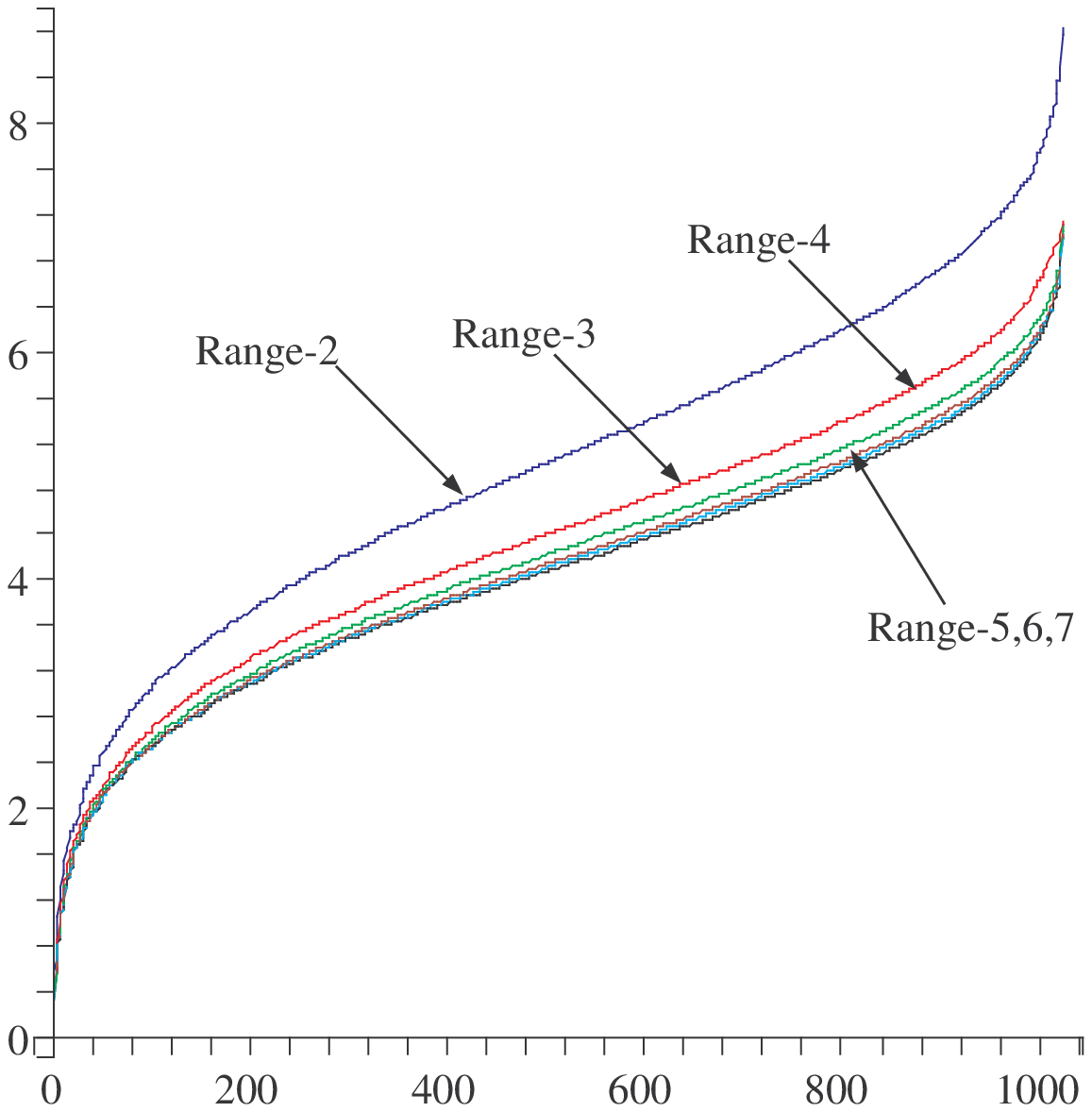,width=3.5in}
\caption{\label{Fig9} 10-Site Comparison }}}
\end{figure}

I begin by defining my blocks to consist of a single lattice site, in which
case the one-block Hamiltonian becomes
\ba
    H_{B_j} &=& \frac{1}{2} p(j)^2 + \frac{\mu^2+2}{2} x(j)^2 \\
    E_0 &=& \frac{1}{2} \sqrt{\mu^2 + 2}
\ea
where $E_0$ is the ground state energy of this system. Since this
is just the Hamiltonian for a simple harmonic oscillator, setting $\mu=0$, we find
the range-$1$ connected term is
\be
    h_1 = \sum_j \left[ \frac{1}{\sqrt{2}} {\bf 1}_j + \sqrt{2}\,\sigma_z(j) \right] .
\ee
Carrying out the $2$-site and $3$-site CORE computations we get the following range-$2$
and range-$3$ connected operators
\ba
    h_2 = \sum_j &&\hskip-15pt\left[ 1.135 {\bf 1}_j + .278\,\sigma_z(j) - .098\,\sigma_x(j)\,\sigma_x(j+1) \right.  \nonumber\\
        &&\left. -.268\, \sigma_y(j)\,\sigma_y(j+1) - .183\,\sigma_z(j)\,\sigma_z(j+1) \right]
\ea
and
\ba
    h_3 = \sum_j &&\hskip-15pt\left[ 1.045 {\bf 1}_j + .138\,\sigma_z(j) - .117\,\sigma_x(j)\,\sigma_x(j+1)
           -.202\, \sigma_y(j)\,\sigma_y(j+1) \right. \nonumber\\
       &&-.167\,\sigma_z(j)\,\sigma_z(j+1)
         +.003\,\sigma_x(j)\,\sigma_x(j+1)\,\sigma_z(j+2) \nonumber\\
         &&-.019\,\sigma_x(j)\,\sigma_z(j+1)\,\sigma_x(j+2) +.003\,\sigma_z(j)\,\sigma_x(j+1)\,\sigma_x(j+2) \nonumber\\
         &&+.032\,\sigma_y(j)\,\sigma_y(j+1)\,\sigma_z(j+2)-.019\,\sigma_y(j)\,\sigma_z(j+1)\,\sigma_y(j+2) \nonumber\\
         &&+.032\,\sigma_z(j)\,\sigma_y(j+1)\,\sigma_y(j+2)
\ea
Clearly the typical coefficients of range-$2$ operators are
significantly larger than the coefficients of range-$3$ operators.
Nevertheless, since as the range of the connected operators grow
more operators appear, it is important to be sure that the effects of the
many small terms don't overwhelm the larger terms in the
renormalized Hamiltonian.  I show that this is not a problem in
Figs.\ref{Fig6}-\ref{Fig9}. To create these plots I computed the
renormalized Hamiltonian out to range-$7$. The
different colored plots exhibit the result of exactly
diagonalizing the Hamiltonian defined by keeping all terms up to
range-$n$.  (What is plotted are the exact eigenvalues in
increasing order versus the integer which labels the eigenvalue after
sorting them in increasing order.) For this reason the last curve in
Fig.\ref{Fig6} and Fig.\ref{Fig7} represents the exact eigenvalues of the
lowest lying $6$-site (or $7$-site) states which have an overlap
with the retained states.  As you can see, after range-$4$, the
longer range terms play only a small role in determining the
eigenvalues over the entire range of energies.  In figures
Fig.\ref{Fig8} and Fig.\ref{Fig9} we see the same plots for up to
$10$-site sublattices. Once again we see that the convergence is
rapid, although I don't bother to plot the corresponding exact
eigenvalues for $8$ or $10$-sites since it would be hard to see
the differences.

The lesson to be learned from these plots is that even in the case
of what would seem to be a very bad approximation to a massless
theory, where the correlation length is infinite, the finite range
cluster approximation to the CORE transformation is rapidly
convergent.

\subsection{Example: Massless Free Fermion}

Now that we have seen that a massive and massless free boson
theory can be mapped into a spin system, with no loss in
information about the low-energy theory, I want to do the same
thing for the case of a massless free
fermion.  In this case, the Hamiltonian is
\ba
    H &=& \sum_{j_1,j_2} \frac{i}{2}\delta'(j_1,-j_2)\psi^\dag_{j_1}\,\sigma_3\,\psi_{j_2} \\
    \delta'(j_1-j_2) &=& -\delta'(j_2-j_1).
\label{hfermi}
\ea
In what follows I will show that fermionic theory maps onto a
highly frustrated anti-ferromagnet, where the fundamental spin
on each site correspond to single-site states of the original
theory having zero-charge\cite{Weinstein:1999hf}.
To do this I expand
$\psi(j)$ in terms of single-site annihilation and creation operators
and define the four possible single-site states to be:
\be
    \ket{0} \quad \ket{+} \quad \ket{-} \quad \ket{\pm} .
\ee
Here, $\ket{0}$, is the state annihilated by the particle and anti-particle destruction
operators.  The states $\ket{+}$ and $\ket{-}$ are one particle and
one anti-particle states (having positive and negative charge), and the
state $\ket{\pm}$ is the zero-charge state containing both a particle
and an anti-particle.  Since the Hamiltonian $H$, Eq.\ref{hfermi}, contains
no terms which refer to a single site, it follows that at the single-site
level these four states are degenerate; therfore
\be
    h_1 = 0 .
\ee
Since there is no reason based on energy considerations to select
any two states to keep, we will define the space of retained states
to be those generated by taking tensor products of the chargeless
single-site states $\ket{0}$ and $\ket{\pm}$.  With this choice we
carry out a CORE transformation to obtain the following
renormalized Hamiltonian
\ba
    h_2 &=& \sum_j \left[ -\frac{1}{4} {\bf 1}_j + \vec{\sigma}(j)\cdot\vec{\sigma}(j+1) \right] \\
    h_3 &=& \sum_j \left[ -.28033 {\bf !}_j + .9428\,\vec{\sigma}(j)\cdot\vec{\sigma}(j+1) +
    .2357\,\vec{\sigma}(j)\cdot\vec{\sigma}(j+2)\right] \\
    h_4 &=& \left[ -.31099 {\bf !}_j + .80001,\vec{\sigma}(j)\cdot\vec{\sigma}(j+1) +
    .23492\,\vec{\sigma}(j)\cdot\vec{\sigma}(j+2)\right. \nonumber\\
    && -.01915\,\vec{\sigma}(j)\cdot\vec{\sigma}(j+3)
    + .03559\,\vec{\sigma}(j)\cdot\vec{\sigma}(j+1)\,\vec{\sigma}(j+2)\cdot\vec{\sigma}(j+3) \nonumber\\
    &&-.08033\,\vec{\sigma}(j)\cdot\vec{\sigma}(j+2)\,\vec{\sigma}(j+1)\cdot\vec{\sigma}(j+3) \nonumber\\
    &&\left. +.0403\,\vec{\sigma}(j)\cdot\vec{\sigma}(j+3)\,\vec{\sigma}(j+1)\cdot\vec{\sigma}(j+2) \right] \\
    H_{\rm ren} &=& \sum_r h_r
\ea
As advertised, we see that the structure of the effective
Hamiltonian is indeed that of a generalized frustrated
anti-ferromagnet and, once again, the cluster expansion is rapidly
convergent.  Moreover, although this computation was done for the
case of a $1+1$ dimensional theory, out to range-$3$ the same
general structure would be obtained for a $1+3$-dimensional
theory.  This strongly implies, at least for special values
of the couplings, that a highly frustrated anti-ferromagnet can exhibit
charge-fractionalization in higher dimensions.  I say this because
the underlying degrees of freedom of this theory are neutral,
however the theory is equivalent to a theory of free fermions.
What other surprises can lurk in this class of theories?

\subsection{The Lattice Schwinger Model}

The lattice Schwinger model is a $1+1$-dimensional gauge theory
defined by the Hamiltonian \be
    H = H_f -\frac{e^2 a}{4} \sum_{m,n} \rho_n\,\vert n-m \vert\,\rho_m ,
\ee
where the charge density operator $\rho_n$ is defined to be
\be
    \rho_n = :\left(\psi^\dag_n\psi_n\right): .
\ee
This model is interesting in that it provides an example of a
confining gauge theory. Moreover, the strong coupling (i.e.,large
$e$) properties of this theory are very similar to the properties
of strong coupling lattice QCD and, in the limit of large $e$,
this problem is very easy to analyze.  Finally, from the point of
view of CORE, although this problem is very difficult to analyze
using earlier real-space renormalization group methods there is no
problem defining CORE transformations starting from locally
gauge-invariant states. This is because, as we saw in the free
fermion theory, it is possible to restrict to locally chargeless
states and still get a non-trivial renormalized Hamiltonian.
In fact, there is even more motivation to truncate to the
single-site chargeless states, $\ket{0}$ and $\ket{\pm}$, which
are the ultimate in locally gauge invariant states, since in this
model they span the set of all low lying physical states for large
$e$.

The space of states spanned by the tensor products of
these chargeless states becomes completely degenerate in the limit
$e\rightarrow \infty$ because they contain no flux.  Thus, it is
possible to expand the Hamiltonian of this system
about the large $e$ limit by doing degenerate perturbation theory
in the kinetic term.  This analysis was carried out in
Ref.\cite{Drell:1976mj}, where it was shown that this strong
coupling theory was equivalent to a frustrated anti-ferromagnet.
The question which dominated this analysis was, ``To what
degree would the strong coupling discussion work as
one moves towards weak coupling?". Kirill Melnikov and I\cite{Melnikov:2000cc}
recently analyzed the lattice Schwinger model and showed that the strong
coupling theory connects smoothly to the weak coupling theory and
one can easily understand all of the properties of the continuum
limit using a generalized form of SLAC fermionic derivative. Space
doesn't permit a full discussion of this issue now, but I mention
it to show that building a CORE truncation procedure using these
locally gauge invariant states is reasonable.

Obviously the strong coupling expansion and CORE computation
agree, however, as one moves away from strong coupling the
coefficients of the range-$r$ terms become non-trivial functions
of $e$.  It is important to point out, however, that the general
structure of the range-$2$, range-$3$ and range-$4$ terms in the
perturbation expansion of the strong coupling theory and
corresponding CORE computation will, for symmetry reasons, be the
same.

Having said this we see that in the strong coupling limit the Schwinger
model is certainly equivalent to a frustrated anti-ferromagnet and moreover,
the CORE computation to range-$3$ says this will be true for all couplings.
Conversely, we see that, at least for special couplings, the general
frustrated anti-ferromagnetic systems can be equivalent either to free massless
fermions or to fermions interacting with gauge-fields.  So now we see that
a frustrated anti-ferromagnet can be equivalent to a theory of free
fermions, or fermions interacting through long-range gauge-fields,
at least for some values of the couplings.

I would now like to finish by talking about QCD.

\section{What About QCD?}

Lattice QCD is much like the lattice Schwinger model in that it is
a theory of quarks interacting with color gauge-fields (defined on
lattice links). Moreover, due to the non-abelian nature of the
gauge-fields, the theory confines at strong coupling, in the same
way fermions are confined in the Schwinger model.  If one uses any
derivative which allows quarks and anti-quarks to live on the same
sites then, as in the Schwinger model, one immediately finds that
at infinite coupling the ground state of the system is highly
degenerate, since all single site color singlet states have
vanishing energy. In other words, at strong coupling quark states
with the quantum numbers of baryons or mesons all have
zero-energy.  This means that if one chooses the space of retained
states to be the one spanned by tensor products of these
zero-energy single-site states, one can then use CORE to construct
an effective theory of interacting mesons and baryons.  If, as in
the Schwinger model, one computes the effective strong coupling
theory\cite{Weinstein:1980jk} to order $1/g^2$ (where $g$ is the gauge
coupling constant) one obtains a generalized, frustrated
anti-ferromagnet whose Hamiltonian takes the form
\be
    H_{\rm eff} =  \frac{1}{g^2} \sum_{j,\mu} Q^{\alpha a}(\vec{j})
    \cdot Q^{\alpha a}(\vec{j}+\vec{\mu})\,S_{\alpha a}(\vec{\mu})
.
\label{hstrong}
\ee
For $n$ flavors of quarks, the operators
$Q^\alpha(\vec{j})$ are the generators of the group $SU(4n)$.
In particular, for the case of $3$ flavors the nearest-neighbor
anti-ferromagnetic interaction leads to a theory with an
$SU(6)\times SU(6)$ chiral symmetry, which breaks spontaneously to
produce a large number of would-be Goldstones bosons. The
next-to-nearest neighbor frustration terms
break this symmetry to chiral $SU(3)\times SU(3)$, which breaks
spontaneously to a theory in which the vector $SU(3)$ symmetry is
exact and the axial-vector symmetry is realized by the existence
of eight Goldstone bosons.

\begin{figure}[2.5in]
\begin{center}
\psfig{figure=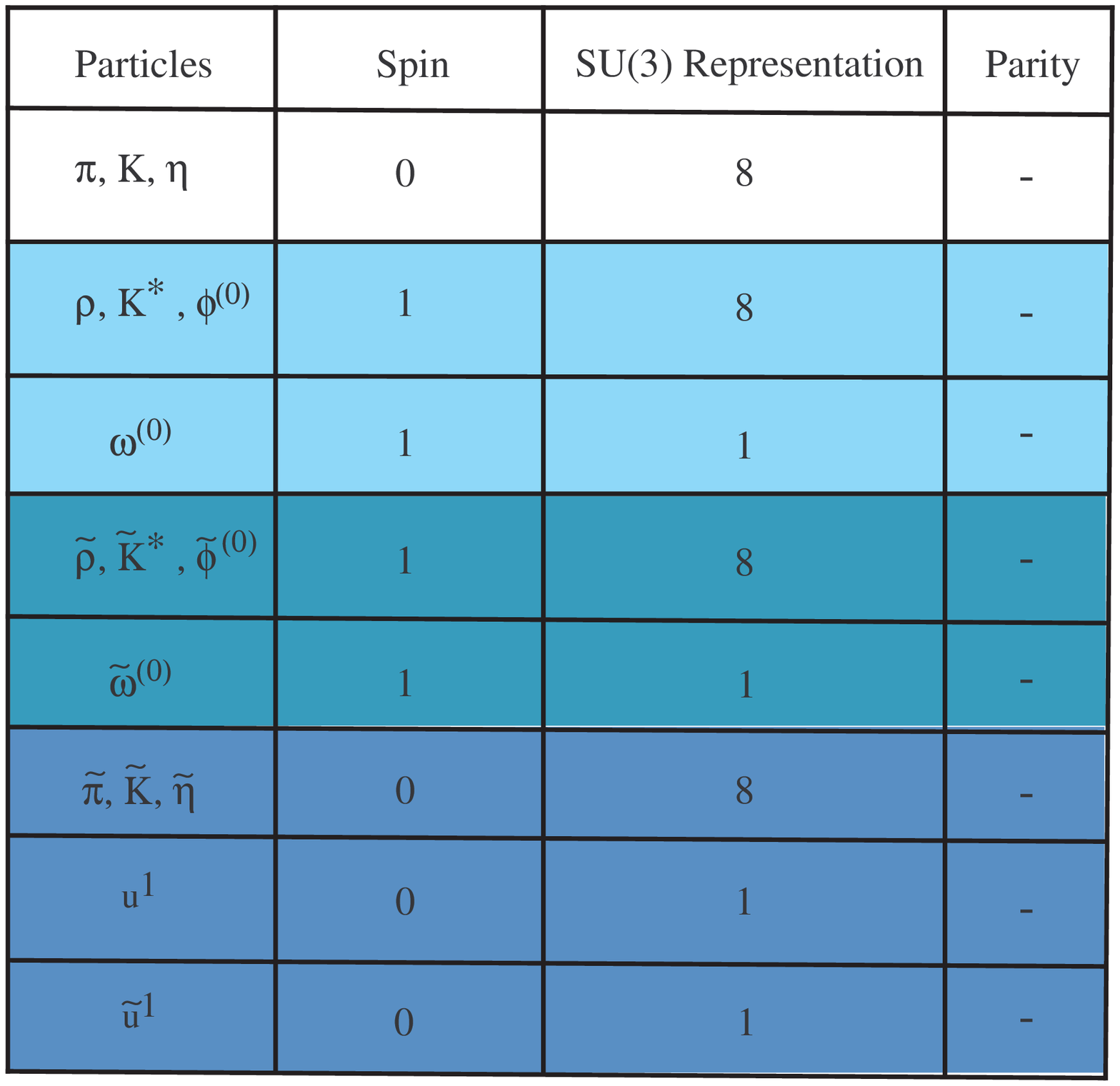,height=3.5in}
\caption{\label{Fig10} Exact and approximate Goldstone bosons when the effects of
next-to-nearest neighbor terms are included.
}
\end{center}
\end{figure}

The explicit form of the next-to-nearest neighbor terms in the
effective Hamiltonian, Eq.\ref{hstrong} allows us to compute the
transformation properties of the $SU(12)$ symmetry breaking terms.
Therefore, it is possible to predict the pattern of
mass-splittings in the large multiplet of would-be Goldstone
bosons without solving the theory exactly.  The table in
Fig.\ref{Fig10} shows the quantum numbers of these particles and
the particles with increasing mass are those shown in darker
shades of blue.  Of course one has to do more than study
the Hamiltonian obtained from strong coupling perturbation theory
to correctly understand the details of the masses, magnetic moments,
etc..  Nevertheless, the observation that out to range-$3$ the effective
Hamiltonian obtained from a CORE computation and that obtained
from strong coupling perturbation theory will be the same, says that
the most important terms in the effective Hamiltonian will have the
same symmetry structure. The difference will be that the values of
the coupling constants appearing in front of the different terms
will now be non-trivial functions of the gauge-coupling $g$.
Assuming, that as in the Schwinger model, the pattern of symmetry
breaking is the same as it is at strong coupling, one obtains
several interesting phenomenological predictions.

First, one obtains the usual Gell-Mann Okubo sum rules for the
mass-splittings in the various $SU(3)$ multiplets.  Second, one
obtains the additional prediction
\be
   m_{k^*}^2 - m_{\rho}^2 = m_L^2 - m_\pi^2
\ee
which is good to about $16\%$, and the prediction
\be
    m_{\tilde{\rho}}^2 \approx 1.2 \pm 0.2 {\rm GeV}
\ee
which is a bit low to match a known state, but after all, this is
just a first order perturbation theory computation.

In addition to these results, the approximate $SU(6)\times SU(6)$
symmetry leads to the famous prediction for the ratio of the
proton to the neutron magnetic moment
\be
   {\mu_p \over \mu_n} = - {3 \over 2} .
\ee
Even more interesting is the fact that, since the weak
axial-vector current is not a generator of the $SU(6)\times SU(6)$
symmetry, the infamous incorrect $SU(6)$ prediction that
\be
    \frac{g_A}{g_V} = - \frac{5}{3}
\ee
is replaced by
\be
    \frac{g_A}{g_V} = - \frac{5}{3}\,X
\ee
where $X$ is some reduced matrix element.  Combining this with the fact that
the usual chiral $SU(3)\times SU(3)$ symmetry is exact, r the
Adler-Weissberger prediction for this ratio should be true.

\section{Conclusions}

To summarize, I hope that I convinced you that CORE techniques can be effectively
used to study Hamiltonian lattice systems.  In addition, I hope to have convinced
you that they can be used to take lattice QCD and map it into an effective
theory whose degrees of freedom only have the quantum numbers of ordinary
mesons and baryons (and glue-balls, if one chooses the fundamental block to be a square).
If one follows Simon Dalley's approach, the usefulness of CORE would be that
this effective theory would be constructed using DLCQ to solve the $1+1$-theory
to high accuracy and construct the space of retained states.  Having done
this one would then couple pairs of lines, etc. and try to evaluate the CORE formula
using similar DLCQ techniques for the coupled system.  In this way
smallish DLCQ computations could be used to construct the an effective
Hamiltonian which contains the low energy physics of QCD.

In any event, even if this approach doesn't appeal to you, I would suggest that
we have at least learned something; namely, that the physics of
generalized frustrated systems is very rich.  This is because, by using CORE to
map known theories into such systems for special values of the coupling constants,
we see that these theories must exhibit a plethora of interesting phases.

\end{document}